\documentclass[aps,prl,twocolumn,showpacs,superscriptaddress,groupedaddress]{revtex4} 
\usepackage{amsfonts}
\usepackage{amsmath}
\usepackage{amssymb}
\usepackage{graphicx}%
\providecommand{\U}[1]{\protect\rule{.1in}{.1in}}

\begin{document}
\title{Trapped-ion Lissajous trajectories}
\author{R. F. Rossetti, G. D. de Moraes Neto, J. Carlos Egues, and M. H. Y. Moussa}
\affiliation{Instituto de F\'{\i}sica de S\~{a}o Carlos, Universidade de S\~{a}o Paulo,
Caixa Postal 369, 13560-970, S\~{a}o Carlos, S\~{a}o Paulo, Brazil}

\begin{abstract}
Here we present a protocol for generating Lissajous curves with a trapped ion
by engineering Rashba- and the Dresselhaus-type spin-orbit interactions in a
Paul trap. The unique anisotropic Rashba $\alpha_{x}$, $\alpha_{y}$ and
Dresselhaus $\beta_{x}$, $\beta_{y}$ couplings afforded by our setup also
enables us to obtain an \textquotedblleft unusual\textquotedblright%
\ Zitterbewegung, i.e., the semiconductor analog of the relativistic trembling
motion of electrons, with cycloidal trajectories in the absence of magnetic
fields. We have also introduced bounded SO interactions, confined to an
upper-bound vibrational subspace of the Fock states, as an additional
mechanism to manipulate the Lissajous {motion} of the trapped ion. Finally, we
accounted for dissipative effects on the vibrational degrees of freedom of the
ion and find that the Lissajous trajectories are still robust and well defined
for realistic parameters.

\end{abstract}

\maketitle

Increasing interest in quantum simulations and controllable trapped ion
systems ---used through the last two decades as a staging platform to
investigate fundamental quantum phenomena \cite{FP} and to
implement quantum information processing \cite{QI}--- have led to the
emulation of the quantum relativistic wave equation. Apart from trapped ions
\cite{Solano,Bermudez}, a number of distinct setups have been used to simulate
zitterbewegung in a variety of physical systems such as quantum wells
\cite{Loss}, photonic crystals \cite{Zhang}, Bose-Einstein condensates
\cite{Spielman} and ultracold atoms \cite{Vaishnav}. It has been demonstrated
that Bose-Einstein condensates can provide an analog to sonic black holes
\cite{Zoller} and the Higgs boson \cite{HB}.

In trapped-ion system Gerritsma \textit{et al}. \cite{Blatt} reported a
proof-of-principle quantum simulation of the one-dimensional Dirac equation
within a trapped-ion experiment. Measuring the time-evolving position of a
single trapped ion set to mathematically simulate a free relativistic quantum
particle, they demonstrated the Dirac Zitterbewegung, as anticipated in Ref.
\cite{Solano}, for distinct initial superpositions of positive- and
negative-energy spinor-like states. The momentum-spin coupling, which
naturally takes place in the relativistic quantum regime, was simulated by the
laser-induced coupling between the vibrational and the electronic states of
the ion.

Motivated by the exciting results in Refs. \cite{Blatt,Solano}, here we
consider a single two-level atom in a two-dimensional Paul trap with four
electrodes, Fig.1, and show how to obtain spin-orbit interactions of the
Rashba and Dresselhaus types, ubiquitous in quantum spintronics and in
emerging areas of condensed matter physics such as topological insulators and
Majorana fermions. Our setup enables us to engineer Hamiltonians with
\textit{anisotropic} SO couplings such as
\begin{equation}
H_{SO}=\frac{\alpha_{x}}{\hbar}\sigma_{x}p_{y}-\frac{\alpha_{y}}{\hbar}%
\sigma_{y}p_{x}+\frac{\beta_{x}}{\hbar}\sigma_{x}p_{x}+\frac{\beta_{y}}{\hbar
}\sigma_{y}p_{y}, \label{Hso}%
\end{equation}
which generalize the canonical form of the SO interactions in semiconductor
quantum wells to include distinct Rashba $\alpha_{x}$, $\alpha_{y}$ and
Dresselhaus $\beta_{x}$, $\beta_{y}$ couplings. In (\ref{Hso}) $\vec{p}$ and
$\vec{\sigma}$ denote the momentum and pseudo-spin operators, respectively, of
the trapped ion. Interestingly, the anisotropic SO couplings can give rise to
physical phenomena not possible in the condensed matter environment, e.g.,
ionic motion with trajectories following Lissajous curves, Fig.1. In Fig. 2 we
show the whole set of Lissajous curves followed by the trapped ion, which we
discuss in detail below. In addition, we can also obtain the \textquotedblleft
unusual\textquotedblright\ Zitterbewegung, a trembling relativistic motion,
with cycloidal trajectories in the absence of magnetic fields \cite{Ze}. We
also demonstrate how to engineer bounded-SO Hamiltonians, confined to an
upper-bound vibrational subspace of the Fock states ---leading to distincts
periodic motions--- apart from analysing the effects of dissipation over
trajectories. 
\begin{figure}[ptb]
\includegraphics[width=0.45\columnwidth]{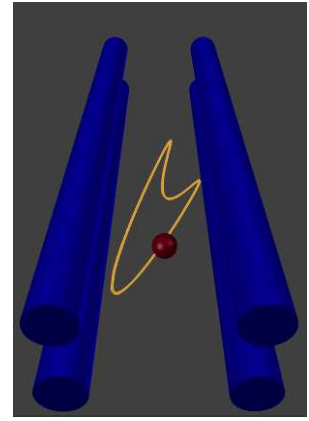}\caption{The schematic of the
experimental Paul trap setup illustrating the ion performing a Lissajous
curve.}%
\label{fig1}%
\end{figure}

\textit{Engineering the anisotropic Rashba interaction.} We
consider a two-level ion acted upon by two laser beams perpendicular to one
another in $x$ and $y$ directions, each providing red and blue sidebands
simultaneously via an electro-optical modulator \cite{Sackett}. From the four
generated frequencies $\omega_{\ell}$ and relative phases $\phi_{\ell}$, two
are tuned to the first-blue sidebands, each coming from one of the original
laser beams. The other two are tuned to the first-red sidebands, leading to
the Hamiltonian ($\hbar=1$)%

\begin{align}
H  &  =i\left[  \Omega_{x}\eta_{x}\left(  e^{i\phi_{1}}a_{x}+e^{i\phi_{2}%
}a_{x}^{\dagger}\right)  \right. \nonumber\\
&  \left.  +\Omega_{y}\eta_{y}\left(  e^{i\phi_{3}}a_{y}+e^{i\phi_{4}}%
a_{y}^{\dagger}\right)  \sigma_{+}+H.c.\right]  \text{,} \label{1}%
\end{align}
where $\eta_{\alpha}$ is the Lamb-Dicke parameter and $\Omega_{\emph{s}}$ the
Rabi frequency for the coupling between the electronic (ground $\left\vert
g\right\rangle $ and excited $\left\vert e\right\rangle $) states of the ion
and its two-dimensional motional degrees of freedom, described by the
annihilation $a_{\emph{s}}$ and creation $a_{\emph{s}}^{\dagger}$ operators
with respect to each direction $\emph{s}=x$,$y$. Moreover, $\sigma
_{+}=\left\vert e\right\rangle \left\langle g\right\vert $ and $\sigma
_{-}=\sigma_{+}^{\dagger}$ are the raising and lowering operators for the
two-level system. By adjusting $\phi_{1}=0,\phi_{2}=\pi,\phi_{3}=\pi/2$, and
$\phi_{4}=3\pi/2$, we obtain, in the interaction picture, the anisotropic 2D
Rashba SO interaction%

\begin{equation}
H=\left(  \Delta_{x}\eta_{x}\Omega_{x}p_{x}\sigma_{y}-\Delta_{y}\eta_{y}%
\Omega_{y}p_{y}\sigma_{x}\right)  \text{,} \label{2}%
\end{equation}
where $p_{\emph{s}}=i\left(  a_{\emph{s}}^{\dagger}-a_{\emph{s}}\right)
/2\Delta_{\emph{s}}$, with $\Delta_{\emph{s}}$ $=1/\sqrt{2m\nu_{\emph{s}}}$,
$m$ being the ion mass and $\nu_{\emph{s}}$ the trap frequency in the
$\emph{s}$ direction. When we adjust the laser parameters so that $\Delta
_{x}=\Delta_{y}$, $\eta_{x}=\eta_{y}$, and $\Omega_{x}=\Omega_{y}$, the above
derived Hamiltonian reduces exactly to the usual Rashba interaction
$H=\gamma\left(  p_{x}\sigma_{y}-p_{y}\sigma_{x}\right)  $ in quantum wells,
where $\gamma=\Delta\eta\Omega$. 

Interestingly, we find that a 3D trap with an
additional laser beam and an appropriate set of phase constants $\phi_{\ell}$
enables us to engineer a 3D Rashba-like term
\begin{equation}
H=\gamma\hat{r}.\left(  \vec{p}\times\vec{\sigma}\right)  \text{,} \label{3}%
\end{equation}
where $\vec{r}$ sets the position of the ion in relation to the trap center.
Unlike the 2D and 3D cases in Eqs. (\ref{2}) and (\ref{3}), the 1D Rashba-like
interaction $H=\gamma p_{x}\sigma_{y}$ does not lead to the unusual
Zitterbewegung with cycloidal orbits. However, when one of the original laser
beams leading to Eq. (\ref{1}) is adjusted to produce a carrier interaction
(instead of the simultaneous red and blue sidebands), we end up with the 1D
interaction%
\begin{equation}
H=\gamma p_{x}\sigma_{y}+\tilde{\Omega}\sigma_{z}\text{.} \label{4}%
\end{equation}
Here, as in the Dirac equation, the energy gap provided by the Stark shift
term $\tilde{\Omega}\sigma_{z}$ ($\tilde{\Omega}$ being also a Rabi
frequency), is a key ingredient for producing Zitterbewegung. In addition, an
appropriate choice of parameters of the laser producing simultaneous blue and
red sidebands leads straightforwardly to a Hamiltonian similar to that in Eq.
(\ref{4}): $H=\eta\Omega\Delta_{x}$ $p_{x}\sigma_{x}+\tilde{\Omega}\sigma_{z}%
$, which simulates the 1D Dirac equation \cite{Solano}.

\textit{Anisotropic Dresselhaus term.} We can also simulate the linear
Dresselhaus SO interaction by adjusting $\phi_{1}=-\phi_{2}=\pi/2$ and
$\phi_{3}=-\phi_{4}=3\pi/2$ in Eq. (\ref{1}); we find
\begin{equation}
H=\gamma\left(  p_{x}\sigma_{x}-p_{y}\sigma_{y}\right)  \text{,} \label{5}%
\end{equation}
which, together with the Rashba coupling, is a crucial ingredient to
manipulate the electron spin in quantum spintronics and, more recently, to
obtain exotic topological phases of matter. We can also consider four laser
beams simultaneously and hence generate a Rashba-Dresselhaus Hamiltonian as
shown in Eq. (\ref{Hso}), with $\alpha_{\emph{s}}=\Delta_{\emph{s}}%
\eta_{\emph{s}}\Omega_{\emph{s}}$ and $\beta_{\emph{s}}=\Delta_{\emph{s}%
}\tilde{\eta}_{\emph{s}}\tilde{\Omega}_{\emph{s}}$, enabling us to simulate
the canonical SO interaction of quantum spintronic in our setup.

At this point it is important to stress that a four-level ion, as in
Ref.~\cite{Solano}, can {also} be used to engineer the interactions of the
forms  [Eqs. (\ref{Hso})-- (\ref{5})]. In this case we have to
consider two additional states $\left\vert g^{\prime}\right\rangle $ and
$\left\vert e^{\prime}\right\rangle $ having the same angular momentum as
$\left\vert g\right\rangle $ and $\left\vert e\right\rangle $, respectively,
but distinct energies and magnetic momenta. For instance, by exciting only the
crossed transitions $g\leftrightarrow e^{\prime}$ and $e\leftrightarrow
g^{\prime}$, we can exactly simulate the mathematical structure of the
Hamiltonian used to obtain the unusual zitterbewegung in Ref. \cite{Ze}. In
this case, the isotropic Rashba-type interaction (\ref{2}) is modified to%

\begin{equation}
H=\tilde{\sigma}_{x}\otimes\Gamma\left(  p_{x}\sigma_{y}-p_{y}\sigma
_{x}\right)  \text{,} \label{6}%
\end{equation}
where $\Gamma=\Delta\eta\Omega$ and $\tilde{\sigma}_{x}$ is the Pauli matrix
in subspace $\left\{  \left\vert g^{\prime}\right\rangle ,\left\vert
e^{\prime}\right\rangle \right\}  $. All other Hamiltonians [Eqs. (\ref{Hso})
-- (\ref{5})] are also modified via a tensor products with $\tilde{\sigma}%
_{x}$. For simplicity, we restrict our analysis to a two-level system which,
as shown below, gives rise to the same dynamics for the unusual Zitterbewegung
as given in Ref. \cite{Ze}.

\textquotedblleft\textit{Bounded\textquotedblright\ spin orbit interaction}.
Before analyzing {the} Lissajous curves in Fig. 2, we would like to address a
case, with no analog in the solid state physics, in which an
\emph{upper-bound} interaction \cite{ToAppear} between the internal and {the}
vibrational degrees of freedom of the ion is engineered. Following the steps
outlined in Ref. \cite{ToAppear}, we can obtain%

\begin{equation}
H_{k=\pm1}=\chi(\eta)\left[  \mathbf{A}(\eta)\sigma_{\pm}+\mathbf{A}^{\dagger
}(\eta)\sigma_{\mp}\right]  \text{,}\label{H}%
\end{equation}
where $\chi(\eta)=\eta\left(  1-\eta^{2}/2\right)  \Omega e^{i\phi}$ and
$\mathbf{A}(\eta)=\left[  1-\eta^{2}a^{\dagger}a/2\right]  a$. Expanding the
operators $A$ and $A^{\dagger}$ in the Fock space basis and adjusting the
Lamb-Dicke parameter to $\eta^{2}=2/N$ $\left[  \eta^{2}=2/\left(  N-1\right)
\text{ with }N\in%
\mathbb{Z}
\right]  $ such that $\mathbf{A}^{\dagger}(\eta)\left\vert N\right\rangle =0$
$\left[  \mathbf{A}(\eta)\left\vert N\right\rangle =0\right]  $, we find that
for an initial vribational state prepared within the \emph{upper-bound
}subspace ranging from $\left\vert 0\right\rangle $ to $\left\vert
N\right\rangle $, the Hamiltonian $H_{k=\pm1}$ becomes%
\begin{equation}
H_{\pm}^{(ub)}=%
{\displaystyle\sum\limits_{n=0}^{N-1}}
\chi_{n}\left(  \left\vert n\right\rangle \left\langle n+1\right\vert
\sigma_{\pm}+H.c.\right)  .\label{HS}%
\end{equation}
where $\chi_{n}=\sqrt{n+1}\left(  1-\eta^{2}n/2\right)  \chi(\eta)$. By using
the Hamiltonian (\ref{Hso}) instead of (\ref{1}) we can derive all the above
spin-orbit interactions but restricted to within the \emph{upper-bound}
subspace. In particular, for the case of the \emph{upper-bound}
Rashba-Dresselhaus interaction, with the same adjustment of the laser
parameters as above we obtain
\begin{equation}
H=\alpha_{x}p_{x}^{ub}\sigma_{y}-\alpha_{y}p_{y}^{ub}\sigma_{x}+\beta_{x}%
p_{x}^{ub}\sigma_{x}-\beta_{y}p_{y}^{ub}\sigma_{y}\text{,}\label{DH}%
\end{equation}
where $p_{\emph{s}}^{ub}=i\left[  \mathbf{A}_{\emph{s}}^{\dagger}%
(\eta)-\mathbf{A}_{\emph{s}}(\eta)\right]  /2\Delta_{\emph{s}}$,
$\alpha_{\emph{s}}=\chi_{n}\eta\Delta_{\emph{s}}$, and $\beta_{\emph{s}%
}=\tilde{\chi}_{n}\eta\Delta_{\emph{s}}$. As should become clearer below, when
discussing the trajectories, the bounded Hamiltonian (\ref{DH}) is able to
generate Lissajous trajectories other than those derived from the usual
Rashba-Dresselhaus interaction. It also enables Lissajous curves from an
initial coherent state, which does not occur with interaction (\ref{Hso}). We
finally note that for each upper-bound parameter $N$ we get distincts\textbf{
}trajectories.

\textit{Cycloidal Zitterbewegung}. Now let us analyze the zitterbewegung with
the (anisotropic) Rashba-like coupling in Eq. (\ref{2}), from which we obtain
the time-dependent components of position operator%
\begin{equation}
\emph{s}(t)=\emph{s}(0)-\frac{c^{2}P_{\emph{s}}}{H}t-iH\left(  c\sigma
_{\emph{r}}+\frac{c^{2}P_{\emph{s}}}{H}\right)  \left(  e^{iHt}-1\right)
\text{,} \label{7}%
\end{equation}
with $\emph{r}\neq\emph{s}=x,y$. We note that Eq. (\ref{7}) exhibits the same
time dependence as that coming from the Dirac zitterbewegung, with the
trembling motion arising from the third term on the right side of the
equality. Hereafter we consider the initial state $\left\vert \psi
(0)\right\rangle =%
{\textstyle\prod\nolimits_{\emph{s}}}
\exp\left[  -(p_{\emph{s}}-p_{0\emph{s}})/2\mu_{\emph{s}}\right]
\otimes\left\vert \varphi\right\rangle $, with both momentum eigenstates
peaked around $p_{0\emph{s}}$ ($\mu_{\emph{s}}$ being the width of the
distribution) and $\left\vert \varphi\right\rangle $ standing for the ionic
internal state, we thus compute the mean value $\left\langle \bar{s}%
(\tau)\right\rangle =\operatorname*{Tr}\left[  \bar{s}\rho(t)\right]  $:
\begin{align}
\left\langle \bar{s}(\tau)\right\rangle  &  =\left\langle \bar{s}%
(0)\right\rangle +\left(  -1\right)  ^{1+\delta_{\emph{s}x}}\varepsilon
^{\delta_{\emph{s}y}}\left\langle \sigma_{\emph{r}}\right\rangle
\tau\nonumber\\
&  +\frac{\varepsilon\bar{p}_{\emph{r}}}{2\xi}\left[  \cos\left(  2\xi
^{1/2}\tau\right)  -1\right]  \left\langle \sigma_{z}\right\rangle \nonumber\\
&  +\frac{\varepsilon\bar{p}_{\emph{r}}}{2\xi^{3/2}}\left[  \sin\left(
2\xi^{1/2}\tau\right)  -2\xi^{1/2}\tau\right] \nonumber\\
&  \times\left(  \left\langle \sigma_{x}\right\rangle \bar{p}_{x}%
+\varepsilon\left\langle \sigma_{y}\right\rangle \bar{p}_{y}\right)  \label{8}%
\end{align}
where we have defined the dimensionless position $\bar{s}=s/\Delta$, momentum
$\bar{p}_{\emph{s}}=2\Delta p_{\emph{s}}$ and time $\tau=\eta\Omega t$, with
$\xi=\sqrt{\left\langle \bar{p}_{\emph{s}}^{2}\right\rangle +\left\langle
\bar{p}_{\emph{r}}^{2}\right\rangle }$ and $\varepsilon=\Delta_{x}^{2}\eta
_{x}\Omega_{x}/\Delta_{y}^{2}\eta_{y}\Omega_{y}$. Here $\delta_{\emph{s}x}$ is
the Kronecker delta and $\varepsilon=1$ ($\neq1$) gives the isotropic
(anisotropic) Rashba-type interaction. We take the momentum state to be a
distribution in the $x$ direction with $\varepsilon^{2}\bar{p}_{y}^{2}\gg
\bar{p}_{x}^{2}$ and the vacuum state in the $y$ direction. With the internal
state $\left\vert \varphi\right\rangle =\left\vert a\right\vert \left\vert
\uparrow\right\rangle +\left\vert b\right\vert \operatorname*{e}%
\nolimits^{i\phi}\left\vert \downarrow\right\rangle $, where $\left\vert
a\right\vert \neq\left\vert b\right\vert $ and $\phi=\pi/2$, such that
$\left\langle \sigma_{x}\right\rangle =0$, we obtain
\begin{subequations}
\begin{align}
\left\langle \bar{x}(\tau)\right\rangle  &  =\left(  \frac{\varepsilon^{2}%
\bar{p}_{y}^{2}}{2\bar{p}_{x}^{2}}-1\right)  \tau-\frac{\varepsilon^{2}\bar
{p}_{y}^{2}}{2\bar{p}_{x}^{3}}\left[  \sin\left(  2\bar{p}_{x}\tau\right)
\right]  \text{,}\label{8a}\\
\left\langle \bar{y}(\tau)\right\rangle  &  =\frac{\varepsilon\bar{p}_{x}%
}{2\xi}\left[  \cos\left(  2\bar{p}_{x}\tau\right)  -1\right]  \text{.}
\label{8b}%
\end{align}
The above equations lead to all the trochoids in Ref. \cite{Ze}, for
appropriate parameters. \begin{figure*}[ptb]
\includegraphics[width=1.0\textwidth]{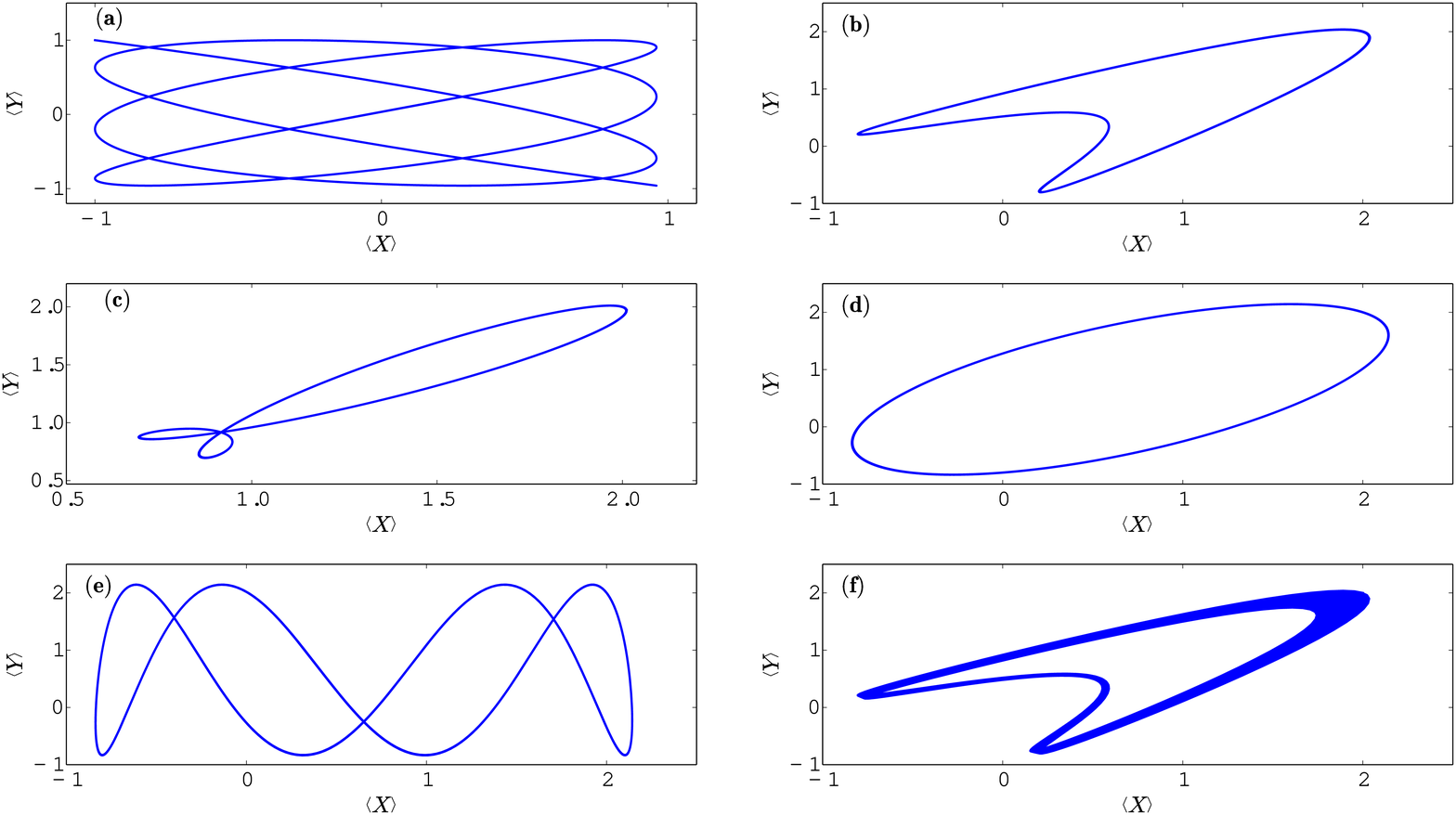}\caption{(a) Lissajou curves
derived from Eq. (\ref{12}) with $\varpi=1.96$ and $\eta\Omega\varpi$ and
$\eta\Omega\sqrt{\varpi}$ around $50$ and $35$ KHz, respectively. (b)
Lissajous curves numerically computed from Eq. (\ref{DH}), starting from
initial coherent state $\Theta=1$ and electronic levels in the superposition
$\left\vert \varphi\right\rangle =\left\vert \uparrow\right\rangle
-i\left\vert \downarrow\right\rangle $, with $N=1$ and $\gamma_{\emph{s}%
}/\Delta_{\emph{s}}=1$. In (c) and (d) we consider the same parameters as in
(b), except for (c)  $\Theta=1-i$ and (d) $N=2$. In (e) we consider the same
initial states as in (b) but $N=2$, $\gamma_{x}/\Delta_{x}=0.4$ and
$\gamma_{y}/\Delta_{y}=1$. Finally, (f) follows from the same parameters as in
(b) but considering dissipative mechanisms in the vibrational degrees of
freedom, with a damping rate $\zeta_{\emph{s}}=10^{-4}\gamma_{\emph{s}}%
/\Delta_{\emph{s}}$.}%
\end{figure*}When considering the Dresselhaus-type interaction (\ref{5}) we
obtain exactly the same time dependence as in Eq. (\ref{8}) but with
$\left\langle \sigma_{x}\right\rangle \bar{p}_{x}+\varepsilon\left\langle
\sigma_{y}\right\rangle \bar{p}_{y}$ changed by $\left\langle \sigma
_{x}\right\rangle \bar{p}_{y}+\varepsilon\left\langle \sigma_{y}\right\rangle
\bar{p}_{x}$, affecting only the amplitude of the curves generated by Eq.
(\ref{8}). This ensures that the cycloidal trajectories without magnetic
fields, derived in Ref. (\cite{Ze}) by considering the effects of an interband
SO term, can also be obtained here from the anisotropic Dresselhaus
interaction. However, when the Rashba- and the Dresselhaus-type interactions
come together as in Eq. (\ref{Hso}), we obtain for the isotropic case
($\alpha_{x}=\alpha_{y}$ and $\beta_{x}=\beta_{y}$)%
\end{subequations}
\begin{align}
\left\langle \bar{s}(\tau)\right\rangle  &  =\left\langle \bar{s}%
(0)\right\rangle +\left[  \left(  -1\right)  ^{1+\delta_{sx}}\left(
\left\langle \sigma_{s}\right\rangle -\kappa\left\langle \sigma_{r}%
\right\rangle \right)  \right]  \tau\nonumber\\
&  -\kappa_{s}\varkappa^{-2}\left\langle \bar{p}_{r}\right\rangle \left\langle
\sigma_{z}\right\rangle \left[  1-\cos\left(  \varkappa\tau\right)  \right]
\nonumber\\
&  -\kappa_{s}\varkappa^{-3}\left\langle \bar{p}_{r}\right\rangle
\mathcal{\Lambda}\left[  \sin\left(  \varkappa\tau\right)  -\tau\right]
\label{9}%
\end{align}
where we have defined the dimensionless coupling $\kappa=\alpha/\beta$, and
the parameters $\kappa_{\emph{s}}=2\left(  \kappa^{2}+\left(  -1\right)
^{1+\delta\emph{s}y}\right)  $, $\varkappa=2\kappa\left[  \kappa\xi
^{2}+2\left\langle \bar{p}_{x}\bar{p}_{y}\right\rangle \right]  $, and
$\Lambda=\left\langle \left(  \kappa\bar{p}_{y}+\bar{p}_{x}\right)  \sigma
_{y}+\left(  \kappa\bar{p}_{x}+\bar{p}_{y}\right)  \sigma_{x}\right\rangle $.
Equation (\ref{9}) shows that with both spin-orbit couplings acting together
with different strengths ($\kappa\neq1$), we still obtain similar trajectories
to those coming from each coupling acting separately. Now, when considering
matched coupling strenghts ($\kappa=1$, i.e., $\alpha=\beta$), two special
situations arise when the electronic levels are prepared as eigenstates of
$\sigma_{z}$ or $\sigma_{\emph{s}}$. In the first case we do not get
Zitterbewegung as expected, since an eigenstate of $\sigma_{z}$ does not
simulate the superposition between positive and negative energy states
required for the Zitterbewegung. However, we do obtain an interesting effect:
we lock the motion of the particle in the $\emph{s}$ direction and obtain an
oscillatory harmonic motion in the $\emph{r}$ direction:
\begin{subequations}
\label{10}%
\begin{align}
\left\langle \bar{s}(\tau)\right\rangle  &  =\left\langle \bar{s}%
(0)\right\rangle -\kappa_{s}\varkappa^{-2}\left\langle \bar{p}_{\emph{r}%
}\right\rangle \text{,}\label{10a}\\
\left\langle \bar{r}(\tau)\right\rangle  &  =\left\langle \bar{r}%
(0)\right\rangle -\kappa_{r}\varkappa^{-2}\left\langle \bar{p}_{s}%
\right\rangle \left[  1-\cos\left(  \varkappa\tau\right)  \right]  \text{.}
\label{10b}%
\end{align}
In the second case we get a uniform motion in the $\emph{s}$ direction and the
expected trembling motion in the $\emph{r}$ direction:
\end{subequations}
\begin{subequations}
\label{11}%
\begin{align}
\left\langle \bar{s}(\tau)\right\rangle  &  =\left\langle \bar{s}%
(0)\right\rangle +\left(  -1\right)  ^{1+\delta_{sx}}\left\langle \sigma
_{s}\right\rangle \tau\text{,}\label{11a}\\
\left\langle \bar{r}(\tau)\right\rangle  &  =\left\langle \bar{r}%
(0)\right\rangle +\left(  -1\right)  ^{1+\delta_{sx}}\tau\nonumber\\
&  -\kappa_{r}\varkappa^{-3}\left\langle \bar{p}_{s}\right\rangle \left\langle
\bar{p}_{y}+\bar{p}_{x}\right\rangle \left[  \sin\left(  \varkappa\tau\right)
-\tau\right]  \text{.} \label{11b}%
\end{align}
Although Eq. (\ref{11b}) does not lead to trajectories similar to those
discussed in Ref. \cite{Ze}, we note that Eq. (\ref{9}) can have its
parameters properly adjusted \ so as to produce cycloidal motion..

\textit{Ionic Lissajous trajectories}. Now we revisit the anisotropic version
of the Rashba-type Hamiltonian (\ref{5}), $H=\left(  \gamma_{x}p_{x}\sigma
_{x}-\gamma_{y}p_{y}\sigma_{y}\right)  $. By adjusting the laser fields such
that $\gamma_{y}^{2}\gg\gamma_{x}^{2}$ and preparing the electronic state as
an eigenstate of $\sigma_{z}$, we obtain
\end{subequations}
\begin{subequations}
\label{12}%
\begin{align}
\left\langle \bar{s}(\tau)\right\rangle  &  =\left\langle \bar{s}%
(0)\right\rangle -\left\langle \bar{p}_{r}\right\rangle \left[  1-\cos\left(
\varpi\tau\right)  \right]  \text{,}\label{12a}\\
\left\langle \bar{r}(\tau)\right\rangle  &  =\left\langle \bar{r}%
(0)\right\rangle +\left\langle \bar{p}_{s}\bar{p}_{r}^{-1}\right\rangle
\left[  1-\cos\left(  \sqrt{\varpi}\tau\right)  \right]  \text{,} \label{12b}%
\end{align}
with $\varpi=\left(  \gamma_{x}/\gamma_{y}\right)  \left\langle \bar{p}%
_{y}\right\rangle $. In Fig. 2(a) we show Lissajous curves governed by Eq.
(\ref{12}) with $\varpi=1.96$ and frequencies $\eta\Omega\varpi$ and
$\eta\Omega\sqrt{\varpi}$around $50$ and $35$ KHz, respectively. In this case,
the Rashba energy $\gamma\sqrt{\bar{p}_{x}+\bar{p}_{x}}$ is around $10^{-10}$
eV. In what follows we obtain Lissajous figures from our previously introduced
bounded SO interaction in Eq. (\ref{DH}), which requires only the preparation
of an initial vibrational coherent state $\left\vert \Theta\right\rangle $
(differently from all the equations of motion derived above, which rely on the
preparation of the initial vibrational state $\left\vert \psi(0)\right\rangle
$). In Fig. 2(b) the mean value of the ionic position has been numerically
computed (running in QuTiP \cite{QuTIP}) from Eq. (\ref{DH}), with $N=1,$
$\gamma_{\emph{s}}/\Delta_{\emph{s}}=1$, and starting from the vibrational
mode in the coherent state $\Theta=1$ and electronic levels in the
superposition $\left\vert \varphi\right\rangle =\left\vert \uparrow
\right\rangle -i\left\vert \downarrow\right\rangle $. In Figs. 2(c) and 2(d)
we consider the same parameters as in Fig. 2(b) except for $\Theta=1-i$ in
Fig. 2(c) and $N=2$ in Fig. 2(d). In Fig. 2(e) we consider $N=2$, $\gamma
_{x}/\Delta_{x}=0.4$, $\gamma_{y}/\Delta_{y}=1$, and the same initial states
as in Fig. 2(b).

\textit{Detrimental effects}. To show that our calculated ionic trajectories
are robust, we have included damping effects due to the environment. Figure 2(f) is similar (same parameters) to in Fig.2(b) but accounts for dissipative
mechanisms in the vibrational degrees of freedom as described by the
Lindbladian $(\zeta_{\emph{s}}/2)\left[  2a_{\emph{s}}\rho a_{\emph{s}%
}^{\dagger}-a_{\emph{s}}^{\dagger}a_{\emph{s}}\rho-\rho a_{\emph{s}}^{\dagger
}a_{\emph{s}}\right]  $, with a damping rate $\zeta_{\emph{s}}=10^{-4}%
\gamma_{\emph{s}}/\Delta_{\emph{s}}$ \cite{exp}. Clearly, the trajectories are
robust and visible for realistic parameters.

We {have presented} a protocol for generating Lissajous curves with the
vibrational motion of an ion in a two-dimensional trap. It relies on the
unique capability of our setup to realize Rashba- and Dresselhaus-type SO
interactions, which allows us to simulate solid-state SO effects within a
highly controllable trapped-ion experiment. We {have} also verified that
Lissajous curves can be derived from upper-bound SO interactions, which may
bring new perspectives to the subject. Addressing some interesting issues to
be investigated further, we first observe that a straightforward extension to
the case of many trapped ions, where the strong and tunable (up to $10^{4}Hz$)
SO strength can be used to explore quantum phase transitions \cite{QPT} and
chaotic behavior \cite{chaos}. Finally, motivated by the results above, we believe it  is worth to investigate the
role of the \textquotedblleft Bounded\textquotedblright\ spin orbit
interaction in solid states systems \cite{s1}.
\end{subequations}
\begin{flushleft}
{\Large \textbf{Acknowledgements}}
\end{flushleft}

The authors acknowledge financial support from PRP/USP within the Research
Support Center Initiative (NAP Q-NANO) and FAPESP, CNPQ and CAPES, the
Brazilian agencies.

\end{document}